\documentstyle[prl,multicol,aps,epsfig]{revtex}

\begin{document}
\title{Soliton electro-optic effects in paraelectrics}
\author{Eugenio DelRe and Mario Tamburrini}
\address{Fondazione Ugo Bordoni, Via B. Castiglione 59, 00142 Roma, Italy,
and INFM, Unita' di Roma I, Italy}
\author{Aharon J. Agranat}
\address{Applied Physics Department, Hebrew University of Jerusalem, Jerusalem 91904, Israel}
\date{\today}
\maketitle

\begin{abstract}
The combination of charge separation induced by the formation of a
single photorefractive screening soliton and an applied external
bias field in a paraelectric is shown to lead to a family of
useful electro-optic guiding patterns and properties.

\end{abstract}

\pacs{42.50.Rh, 42.65.Pc, 03.65.Kf}

\begin{multicols}{2}
\narrowtext Apart from their inherent interest as peculiar
products of nonlinearity, spatial solitons hold the promise of
allowing viable optical steering in bulk environments
\cite{phystoday} \cite{gricio}. Photorefractive screening solitons
differ from other known manifestations of spatial self-trapping
for their peculiar ease of observation and versatility
\cite{bruno}, and recent experiments in photorefractive
strontium-barium-niobate (SBN) and potassium-niobate (KNbO$_3$)
have demonstrated two conceptual applications of their guiding
properties. In the first case, a tunable directional coupler was
realized making use of two independent slab-solitons
\cite{dircoupl}; in the second, self-induced phase-matching was
observed to enhance second-harmonic-generation \cite{shg}.
Although results suggest a means of obtaining all-optical
functionality, actual implementation is hampered by the generally
slow nonlinear response \cite{solymar}, that can be "accelerated"
only at the expense of stringent intensity
requirements\cite{motipulsed}. In contrast, non-dynamic guiding
structures have been observed by fixing a screening soliton
\cite{fixing}, or in relation to the observation of spontaneous
self-trapping during a structural crystal phase-transition
\cite{spont}. One possible method of obtaining acceptable dynamics
is to make directly use of the electro-optic properties of the
ferroelectrics involved, in combination with the internal
photorefractive space charge field deposited by the soliton. Since
photorefractive charge-activation is wavelength dependent, one can
induce charge separation in soliton-like structures at one active
wavelength (typically visible), and then read the electro-optic
index modulation at a different, nonphotorefractive, wavelength
(typically infrared) \cite{waveguiding} \cite{linear}.  For {\it
noncentrosymmetric} samples (such as the above mentioned crystals)
that typically host screening soliton formation, the electro-optic
index of refraction modulation is proportional to the static
crystal polarization {\bf P}, and thus to the electric field
(linear electro-optic effect).  For these, {\it no} electro-optic
modulation effects are possible: for whatever value of external
constant electric field E$_{ext}$, the original soliton supporting
guiding pattern remains {\it unchanged}.  In {\it
centrosymmetrics}, such as photorefractive
potassium-lithium-tantalate-niobate (KLTN), solitons are supported
by the quadratic electro-optic effect \cite{motironnie} \cite{eu1}
\cite{eu2} \cite{ronnieyariv}. In this case, the "nonlinear"
combination of the internal photorefractive field with an external
electric field can give rise to new and useful soliton-based
electro-optic phenomena, which we here study for the first time.

The basic mechanism leading to screening soliton formation is the
following: a highly diffracting optical beam ionizes impurities
hosted in the lattice of an electro-optic crystal.  An externally
applied electric field makes these mobile charges drift to less
illuminated regions, forming a double layer that renders the
resultant electric field in the illuminated region lower.  For an
appropriate electro-optic sample, this leads to a self-lensing and
soliton propagation, when beam diffraction is exactly compensated.
For slab solitons, i.e. those self-trapped beams that originate
from a beam that linearly diffracts only in one transverse
dimension (x), for a given soliton intensity
full-width-half-maximum (FWHM) $\Delta$x, a given ratio between
the soliton peak intensity and the (generally artificial)
background illumination, $u_0^2=I_{peak}/I_b$ (intensity ratio),
solitons form for a particular value of applied external biasing
field $\overline{E}$. The soliton-supporting electric field E is
expressed by E=(V/L)(1+I(x)/I$_b$)$^{-1}$, where V is the external
applied voltage, L is the distance between the crystal electrodes
(thus $\overline{E}$=V/L), and I(x) is the soliton optical
intensity confined in the x transverse dimension
\cite{motironnie}. This electric field, a result of a complex
nonlinear light-matter interaction, is present even when the
generating optical field is blocked, and the sample is illuminated
with a nonphotorefractively active light. Charge separation is
smeared out only by slow recombination, associated with dark
conductivity, characterized by considerably long decay times. The
nonphotorefractively active illumination, although not leading to
any further evolution in the internal charge field, will feel the
index inhomogeneity due to the quadratic electro-optic response
described by the relation $\Delta$n =-(1/2) n$^3$ g$_{eff}
\epsilon_0^2 $($ \epsilon_r$-1)$^2$E$^2$, where n is the crystal
index of refraction, $g_{eff}$ is the effective electro-optic
coefficient for a given scalar configuration, $\epsilon_0$ is the
vacuum dielectric constant, and $\epsilon_r$ is the relative
dielectric constant.  The actual electric field in the crystal is
now E=(V/L)(1+I(x)/I$_b$)$^{-1}$-(V/L)+E$_{ext}$, where E$_{ext}$
(in general $\neq \overline{E}$) is the externally applied
electric field {\it after} the nonlinear processes have occurred
(the "read-out" field).  The index pattern induced is

\begin{equation}
\label{indexreconstruction} \Delta n =-\Delta n_0
\left(\frac{1}{1+I(x)/I_b}-1+\frac{E_{ext}}{V/L} \right)^2,
\end{equation}
where $\Delta$n$_0$ =(1/2) n$^3$g$_{eff} \epsilon_0^2 $($
\epsilon_r$-1)$^2$(V/L)$^2$.  In Fig.(1) we show two families of
induced index patterns associated with two solitons at different
saturation levels.  In Fig.(1a) a 7$\mu$m FWHM soliton at
$\lambda$=514 nm wavelength ($\Delta$n$_0\simeq$5.4$\times
10^{-4}$, for n=2.45) with an intensity ratio $u_0^2$=4, leads to
three characteristic pattern regimes: for $\eta =
$E$_{ext}/(V/L)\simeq$1, the soliton supporting potential is
formed. For $\eta \simeq$0, an antiguiding hump appears, whereas
for intermediate values of $\eta$, a twin-waveguide potential
forms. Analogous results can be predicted for a strongly saturated
regime shown in Fig.(1b), where a  11$\mu$m soliton is formed for
$u_0^2\simeq$22.

\begin{figure}
\begin{center}
\resizebox{8.5cm}{!}{\mbox{\includegraphics*[3cm,6cm][27cm,15.5cm]{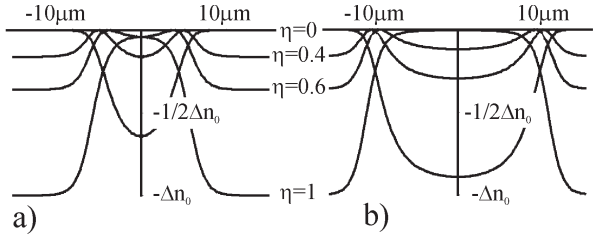}}}
\caption{Predicted electro-optic index patterns resulting from the
soliton deposited space-charge field, for $u_0=2$ (a) and
$u_0=4.7$ (b).} \label{potentials}
\end{center}
\end{figure}

Experiments are carried out with an apparatus that is well
documented in literature \cite{eu1} \cite{eu2}.  An enlarged
TEM$_{00}$ Gaussian beam from a CW Argon-ion laser operating at
$\lambda$ =514nm, is focused be means of an f=150mm cylindrical
lens onto the input facet of an
$3.7^{(x)}\times4.6^{(y)}\times2.4^{(z)}$ mm sample of zero-cut
paraelectric KLTN, at T=20 $^\circ$C (with a critical temperature
T$_c$=11 $^\circ$C), giving rise to an approximately
one-dimensional x-polarized Gaussian beam of $\Delta$x$\cong$11
$\mu$m ("soliton" beam), and the entire crystal is illuminated
with a second, homogeneous beam ("background" beam) from the same
laser, polarized along the y axis.  Both the focused and the
plane-wave beams copropagate along the z-direction. The constant
voltage V is applied along the crystal x direction, the crystal
itself being doped with Vanadium and Copper impurities, and
photorefractively active at the laser wavelength. Guiding patterns
can be investigated either by illuminating the crystal with an
infrared beam (as mentioned above), or simply by using the same
soliton-forming wavelength, but at a lower intensity, since
photorefractive temporal dynamics are proportional to beam
intensity. Here we use this read-out method, and in what follows
all read/write experiments are at $\lambda$=514nm, with
I$_{read}$/I$_{write}\cong$20. By changing the value of the
applied readout voltage, V$_{ext}$, we can explore the optical
potential described by Eq.(1), through the variable $\eta$. Beam
distribution is investigated by imaging the facets of the sample
onto a CCD camera by means of a second lens placed after the
sample (along the z direction).

\begin{figure}
\begin{center}
\resizebox{8.5cm}{!}{\mbox{\includegraphics*[9cm,8cm][20cm,13.5cm]{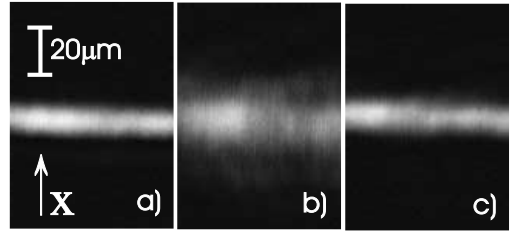}}}
\caption{Soliton formation: an input 11$\mu$m beam (a) diffracts
to 24 $\mu$m in linear propagation (V=0) (b) and self-traps for
V$_{exp}$=1.33 kV at T=20$^\circ$C, for $u_0\simeq$4.7.}
\label{soliton}
\end{center}
\end{figure}

In Fig.(2) the observation of a single photorefractive screening
soliton is shown.  The 11$\mu$m soliton is observed with an
intensity ratio u$_0^2\cong$ 22 at V$_{exp}$=1.33 kV, annulling
linear diffraction to 24 $\mu$m .  Soliton formation takes
approximately 3 min, for an I$_{peak}\simeq$1.8 kW/m$^2$ (I$_b
\simeq$80 W/m$^2$), measured directly before the sample, thus
meaning that erasure during readout would take, at the very least,
about 1 hr (i.e. longer than the duration of any one of our
experiments).  Had we used an IR read-out beam, decay would be
halted indefinetly. Given the sample g$_{eff}$=0.12m$^4$C$^{-2}$,
$\epsilon_r \simeq$9000, $\Delta n_0 \simeq$ 6.9$\times10^{-4}$,
the expected value for soliton formation would be
V$_{th}\simeq$1.27 kV.

\begin{figure}
\begin{center}
\resizebox{8.5cm}{!}{\mbox{\includegraphics*[7cm,8cm][22cm,15.5cm]{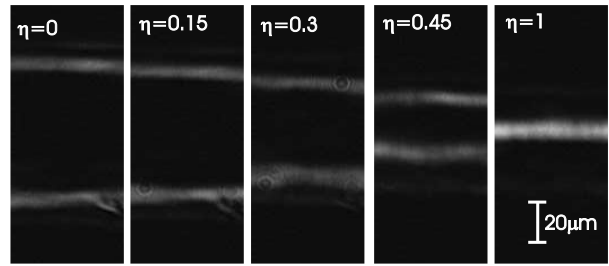}}}
\caption{Output light distribution of the read-out beam.  For
$\eta$=0-0.3 the beam is scattered.  For $\eta$=0.45 the twin beam
structure forms, whereas for $\eta$=1 the original guiding pattern
emerges.} \label{electro-optic}
\end{center}
\end{figure}

In Fig.(3) we show the same region of the crystal invested by the
less intense (but otherwise identical to the soliton generating)
"read" beam at various values of $\eta$. For $\eta$=1 the output
beam is identical to the soliton (apart from the actual
intensity). For low values of $\eta$ ($\eta<$0.4) the index
pattern given by Eq.(1) is antiguiding, and the output beam is
scattered and split into two diffracting beams (beam "bursting",
see Fig.(1b)). As $\eta$ is increased, the defocusing is weakened
and for $\eta\approx0.45$ the sample gives rise to a
beam-splitting on the twin-waveguide structure formed by the
two-hump potential, as shown in Fig.(1). The distance between the
two beams is $\approx 20 \mu$m.  As opposed to previous
defocusing, in this case light is exciting a guided mode.

\begin{figure}
\begin{center}
\resizebox{8.5cm}{!}{\mbox{\includegraphics*[8cm,8cm][22cm,12cm]{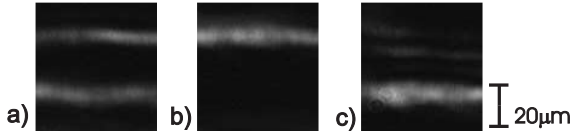}}}
\caption{Electro-optic switching.  The output light distribution
of the read beam (a) for $\eta$=0.45, the side-guided beam, when
the crystal is shifted sideways of 10 $\mu$ m (b), the output in
the same condition, but with $\eta$=0.8 (c).} \label{switching}
\end{center}
\end{figure}

Next we shift the crystal with respect to the optical beam in the
x direction, so as to launch it directly into one of the
twin-guides for intermediate values of $\eta$. For an $\eta$=
0.45, shifting the crystal by 10$\mu$m, the beam is guided by the
side hump, as shown in Fig.(4b).  In this forward guiding
condition, we change $\eta$ to $\eta$=0.8.  The potential commutes
from a double-hump twin-waveguide to a single guiding pattern (see
Fig.1). The optical beam is redirected as shown in Fig.(4c).

It is therefore possible to realize, by means of the formation of
a single photorefractive centrosymmetric screening soliton, three
qualitatively different optical circuits: a single waveguide, a
double waveguide beam-splitter, and an antiguiding beam-stopper.
If the crystal is shifted so as to launch the guided beam into one
of the twin-guides, it is possible to deviate the beam,
maintaining its strong confinement, realizing an electro-optic
switch.  Had we used a longer sample, launching the beam in a
twin-waveguide leads to a tunable directional coupler, as shown in
Fig.(5).

The observed phenomena represent an important step in the
achievement of viable soliton based components in two major
aspects.  The first is that the observed phenomenology occurs with
the formation of a single soliton, that is only used to deposit a
pattern of charge displacement (a peculiar volume hologram),
whereas switching from one regime to the other occurs only through
the change of the applied electric field.  Thus switching dynamics
are only limited by capacity charging times, as all other
electro-optic devices.  Secondly, whereas screening soliton {\it
formation} requires a constant applied external field, during
read-out, the use of independent electrodes can allow the
formation of composite circuitry in cascade, all from a single
soliton.


\begin{figure}
\begin{center}
\resizebox{8.5cm}{!}{\mbox{\includegraphics*[4cm,6cm][27.4cm,13cm]{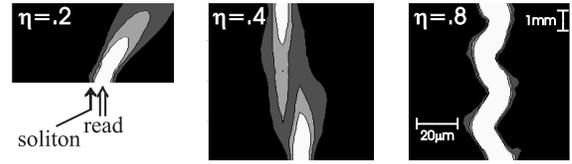}}}
\caption{Predicted evolution of an $\approx$ 7$\mu$m beam: top
view of read-out in an 8mm sample for $\eta$=0.2 (beam deflection
and diffraction) (left), one beat directional coupling for $\eta$
=0.4 from right hump to left hump (center), and mode beating for
$\eta$ =0.8 ($\approx$ 2mm mode beat).} \label{mode-coupling}
\end{center}
\end{figure}

The work of E.D. and M.T. was carried out in the framework of an
agreement between Fondazione Ugo Bordoni and the Italian
Communications Administration. Research carried out by A.J.A. is
supported by a grant from the Ministry of Science  of the State of
Israel.

\end{multicols}

\end{document}